\documentclass[12pt]{article}
\usepackage{amssymb}

\usepackage[dvips]{graphicx}
\usepackage{epsfig}
\usepackage{amsmath}
\usepackage{float}

\newcommand{\bm}{\begin{multiline}}
\newcommand{\beq}{\begin{equation}}
\newcommand{\eeq}{\end{equation}}
\newcommand{\beqs}{\begin{eqnarray}}
\newcommand{\eeqs}{\end{eqnarray}}

\begin{document}

\thispagestyle{empty}

\begin{flushright}
hep-th/0608037\\
\end{flushright}

\hfill{}

\hfill{}

\hfill{}

\vspace{32pt}

\begin{center}
\textbf{\Large Note on counterterms in asymptotically flat spacetimes} \\[0pt]

\vspace{48pt}

\textbf{Dumitru Astefanesei,}\footnote{E-mail: \texttt{dastef@hri.res.in}} \textbf{Robert B. Mann}\footnote{
E-mail: \texttt{mann@sciborg.uwaterloo.ca}} \textbf{and Cristian Stelea}\footnote{E-mail: \texttt{cistelea@uwaterloo.ca}}

\vspace*{0.2cm}

\textit{$^{1}$Harish-Chandra Research Institute}\\[0pt]
\textit{Chhatnag Road, Jhusi, Allahabad 211019, India}\\[.5em]

\textit{$^{2}$Perimeter Institute for Theoretical Physics}\\[0pt]
\textit{31 Caroline St. N. Waterloo, Ontario N2L 2Y5 , Canada}\\[.5em]

\textit{$^{2,3}$Department of Physics, University of Waterloo}\\[0pt]
\textit{200 University Avenue West, Waterloo, Ontario N2L 3G1, Canada}\\[.5em%
]
\end{center}

\vspace{30pt}

\begin{abstract}
We consider in more detail the covariant counterterm proposed by Mann and
Marolf in asymptotically flat spacetimes. With an eye to specific practical
computations using this counterterm, we present explicit expressions in
general $d$ dimensions that can be used in the so-called `cylindrical
cut-off' to compute the action and the associated conserved quantities for
an asymptotically flat spacetime. As applications, we show how to compute
the action and the conserved quantities for the NUT-charged spacetime and
for the Kerr black hole in four dimensions.
\end{abstract}

\setcounter{footnote}{0}



\section{Introduction}

Over the years many expressions have been proposed for computing conserved
quantities in asymptotically flat spacetimes. The general idea in such
constructions is to study the asymptotic values of the gravitational field,
far away from an isolated object, and compare them with those corresponding
to a gravitational field in the absence of the respective object \cite{Wald}%
. However, most of these proposals will provide results that are relative to
the choice of a reference background --- they are usually called background
subtraction methods. The background must be chosen such that its topological
properties match the solution whose action and conserved charges we want to
compute. However, this does not unequivocally fix the choice of the
background \cite{CCM} and moreover, there might be cases in which the
topological properties of the solution rule out any natural choice of the
background.

Inspired by the AdS/CFT correspondence, an alternative procedure ---
referred to as the `counterterm method' --- has been proposed \cite{Kraus,
Lau, Mann1}. In this approach one supplements the action by including
suitable boundary counterterms. These counterterms are functionals only of
curvature invariants of the induced metric on the boundary and so they do
not affect the equations of motion. By choosing appropriate counterterms
that cancel the divergences, one can then obtain \textit{finite} expressions
for the action. Unlike the background subtraction methods, this counterterm
procedure is intrinsic to the spacetime of interest and it is unambiguous
once the counterterm action is specified. However, while there is a general
algorithm for generating the counterterms for asymptotically AdS spacetimes %
\cite{Kraus}, the asymptotically flat case has been considerably
less-explored. Early proposals \cite{Lau,Mann1,Ho} engendered study of
proposed counterterm expressions for a class of $d$-dimensional
asymptotically flat solutions whose boundary topology is $S^{n}\times
R^{d-1-n}$ \cite{Kraus}.

Recently there have been intresting developments in this area. Astefanesei
and Radu proposed a \textit{renormalized} stress-tensor for a particular
family of locally asymptotically flat spacetimes \cite{Astefanesei:2005ad}
--- it was computed by varying the total action (including the counterterms)
with respect to the boundary metric. Conserved quantities can be constructed
from this stress-tensor via the algorithm of Brown and York \cite%
{Brown:1992br}. Subsequently, Mann and Marolf have generalized this method
to arbitrary asymptotically locally flat spacetimes \cite{Mann:2005yr}. By
using a new local, covariant counterterm they obtained conserved quantities
that agree with older definitions known in literature. In particular, they
constructed a boundary stress-tensor similar with the one used in
quasi-local context \cite{Brown:1992br,CCM}, which in the so-called
`hyperbolic cut-off' has been shown to lead to conserved quantities that are
built out of the electric part of the Weyl tensor, similar to the
Ashtekar-Hansen expressions in four dimensions \cite{AshtekarHansen}. The
connection with the Ashtekar-Hansen conserved quantities has been examined
in more detail in \cite{Mann:2006bd}, in which it was argued that the
canonical form of the action built using the Mann-Marolf (MM) counterterm
reduces to the ADM action \cite{ADM}. Hence the conserved quantities should
agree with the ADM conserved charges.

This method was also used in computing the gravitational energy of the
Kaluza Klein monopole \cite{KKenergy} and of $(2+1)$ Minkowski space \cite%
{Marolf:2006xj}, the mass of Kaluza-Klein black holes with squashed horizons %
\cite{Cai:2006td,eugen1,wang,Yazadjiev:2006iv}, and the mass of non-uniform black string
solutions \cite{Kleihaus:2006ee}.

The main motivation for the present work is to investigate more closely the
local MM counterterm prescription for computing the action and conserved
charges associated with asymptotically flat spaces. We will specifically
focus on the so-called `cylindrical cut-offs' of the spacetime. We exhibit
explicit expressions for the boundary stress-tensor and we show how to
compute the general renormalised action in $d$-dimensions. As an example of
this counterterm technique, we compute the action and the conserved
quantities for NUT-charged spaces in four dimensions. Using also a distinct
proposal for the boundary counterterm, namely the counterterm proposed by
Lau \cite{Lau} and Mann \cite{Mann1} we find agreement in both cases with
previous results in literature. We consider also the conserved charges and
action for a rotating black hole in four dimensions.

The structure of this paper is as follows: in the next section we review the
MM counterterm prescription and show how to compute the boundary
stress-tensor and action in cylindrical cut-offs. In Section $3$ we briefly
review other counterterm proposals for asymptotically flat spacetimes, while
in Section $4$ we apply them to computation of the action and conserved
quantities for NUT-charged spacetimes. In Section $5$ we compute the action
and conserved charges for the Kerr solution. The last section is dedicated
to conclusions.

\section{The Mann-Marolf counterterm}

For asymptotically flat spacetimes, the gravitational action consists of the
bulk Einstein-Hilbert term, supplemented by the boundary Gibbons-Hawking
term in order to have a well-defined variational principle \cite%
{Gibbons:1976ue}. In general $d$-dimensions, the gravitational action for an
asymptotically flat spacetime is then taken to be: 
\begin{equation}
I_{B}+I_{\partial B}=-\frac{1}{16\pi G}\int_{M}d^{d}x\sqrt{-g}R-\frac{1}{%
8\pi G}\int_{\partial M}d^{d-1}x\sqrt{-h}K.  \label{actbulkgh}
\end{equation}%
Here $M$ is a $d$-dimensional manifold with metric $g_{\mu \nu }$, $K$ is
the trace of the extrinsic curvature $K_{ij}=\frac{1}{2}h_{i}^{k}\nabla
_{k}n_{j}$ of the boundary $\partial M$ with unit normal $n^{i}$ and induced
metric $h_{ij}$.

When evaluated on non-compact solutions of the field equations, it turns out
that both terms in (\ref{actbulkgh}) diverge. The general remedy for this
situation is to add a counterterm, \textit{i.e.} a coordinate invariant
functional of the intrinsic boundary geometry that is specifically designed
to cancel out the divergencies.

In \cite{Mann:2005yr}, Mann and Marolf put forward a new counterterm that is
also given by a local function of the boundary metric and its curvature
tensor. The new counterterm is taken to be the trace $\hat{K}$\ of a
symmetric tensor $\hat{K}_{ij}$ that is defined implicitly in terms of the
Ricci tensor $\mathcal{R}_{ij}$ of the induced metric on the boundary via
the relation 
\begin{equation}
\mathcal{R}_{ik}=\hat{K}_{ik}\hat{K}-\hat{K}_{i}^{m}\hat{K}_{mk}.
\label{kmm}
\end{equation}
In solving (\ref{kmm}), one chooses a solution $\hat{K}_{ij}$ that
asymptotes to the extrinsic curvature of the boundary of Minkowski space as $%
\partial M$ is taken to infinity. Therefore, in contrast to previous
counterterm proposals this new counterterm assigns an identically zero
action to the flat background in any coordinate systems, while giving finite
values for asymptotically flat backgrounds. The renormalized action leads to
the usual conserved quantities that can also be expressed in terms of a
boundary stress-tensor whose leading-order expression involves the electric
part of the Weyl tensor: 
\begin{equation}
T_{ij}=\frac{1}{8\pi G}\frac{\rho E_{ij}}{d-3}+\mathcal{O}(\rho^{-(d-3)}).
\label{Thyp}
\end{equation}%
Here $E_{ij}$ is the pull-back to the boundary of the contraction of the
bulk Weyl tensor $C_{\mu\nu\rho\tau}$ with the induced metric $h^{\mu\nu}$.
More precisely, introducing the unit normal vector $n^{\mu}$ to the boundary 
$\partial M$ then the electric part of the bulk Weyl tensor is defined by

\begin{equation*}
E_{\mu \nu }=C_{\mu \rho \nu \tau }n^{\rho }n^{\tau }=-C_{\mu \rho \nu \tau
}h^{\rho \tau },
\end{equation*}%
and $E_{ij}$ is simply the pull-back to the boundary of the above tensor.
Notice that we are using here the so-called `hyperbolic cut-off', in which
the line-element is taken to admit the following asymptotic expansion at
spatial infinity: 
\begin{eqnarray}
ds^{2} &=&\left( 1+\frac{2\sigma }{\rho ^{d-3}}+\mathcal{O}(\rho
^{-(d-2)})\right) d\rho ^{2}+\rho ^{2}\left( h_{ij}^{0}+\frac{h_{ij}^{1}}{%
\rho ^{d-3}}+\mathcal{O}(\rho ^{-(d-2)})\right) d\eta ^{i}d\eta ^{j}  \notag
\\
&&+\rho \left( \mathcal{O}(\rho ^{-(d-2)})\right) d\rho d\eta ^{j}.
\label{hypcut}
\end{eqnarray}

Here $h_{ij}^0$ and $\eta^i$ are a metric and the associated coordinates on
the unit $(d-2,1)$ hyperboloid $\mathcal{H}^{d-1}$, while $\sigma$ and $%
h_{ij}^1$ are smooth tensorial fields defined on $\mathcal{H}^{d-1}$.
Moreover, $\rho$ plays here the role of a `radial' coordinate such that
spacelike infinity is reached in the $\rho\rightarrow\infty$ limit with
fixed $\eta$ and the symbols $\mathcal{O}(\rho^{-(d-2)})$ refer to terms
that fall-off at least as fast as $\rho^{-(d-2)}$ in the $%
\rho\rightarrow\infty$ limit.

While the hyperbolic cut-off is more natural for covariant investigations,
its use in practical applications is cumbersome since in order to apply the
above results one has to transform every solution into a form that is
asymptotically of the form (\ref{hypcut}). Such coordinate transformations
can be very involved. Moreover, even if the conserved charges can be found
by using the boundary stress-tensor to leading order, to evaluate the finite
regularised action one has to find the next-to-leading terms in $\hat{K}_{ij}
$ \cite{Mann:2006bd}. For these reasons, it is desirable to find directly
the corresponding expressions for the regularised action and conserved
charges in the `cylindrical cut-offs'. In the remainder of this section we
will focus on the so-called `cylindrical cut-off', in which the asymptotic
form of the metric at spatial infinity takes the form 
\begin{eqnarray}
ds^{2} &=&-\left( 1+\mathcal{O}(\rho ^{-(d-3)}\right) )dt^{2}+\left( 1+%
\mathcal{O}(\rho ^{-(d-3)}\right) )dr^{2}+r^{2}\left( \omega _{IJ}+\mathcal{O%
}(\rho ^{-(d-3)}\right) )d\theta ^{I}d\theta ^{J}  \notag \\
&&+\mathcal{O}(\rho ^{-(d-4)})d\theta ^{I}dt,  \label{cycut}
\end{eqnarray}%
where $\omega _{IJ}$ and $\theta ^{I}$ are the metric and coordinates on the
unit $(d-2)$-sphere.

For vacuum metrics the Einstein-Hilbert action vanishes on solutions. We are
therefore interested in computing the boundary terms only%
\begin{equation*}
I_{\partial B}=-\frac{1}{8\pi G}\int_{\partial M}d^{d-1}x\sqrt{-h}(K-\hat{K}%
).
\end{equation*}%
where we note that the extrinsic curvature satisfies the Gauss-Codazzi
relation 
\begin{equation*}
\mathcal{R}_{ik}=R_{ikjl}h^{kl}+K_{ik}K-K_{i}^{m}K_{mk},
\end{equation*}%
which is very similar to (\ref{kmm}). The only difference is the presence of
the term $R_{ikjl}h^{kl}$, where $R_{ikjl}$ is the pullback to the boundary $%
\partial M$ of the bulk Riemann tensor. The leading terms of $K_{ij}$ and $%
\hat{K}_{ij}$ are given by the extrinsic curvature of the standard cylinder
of radius $r$ in Minkowski spacetime:

\begin{eqnarray}
K_{ij}\sim\hat{K}_{ij}&=&r\omega_{ij}+\mathcal{O}(\rho^{-(d-4)}).
\end{eqnarray}

Here, $\omega _{ij}$ is the pull-back to $\partial M$ of the round metric $%
\omega _{IJ}$ on the unit sphere $S^{d-2}$.  Furthermore, one has to solve
the linearised Gauss-Codazzi relation: 
\begin{equation}
R_{ikjl}h^{kl}=-\Delta _{kl}h^{kl}\hat{K}_{ij}-\Delta _{ij}\hat{K}+\Delta
_{il}\hat{K}_{j}^{l~}+\Delta _{jl}\hat{K}_{i}^{l~}  \label{LGC}
\end{equation}%
to compute the difference $\Delta _{ij}\equiv K_{ij}-\hat{K}_{ij}$ to first
order \cite{Mann:2005yr,Mann:2006bd}. It is convenient at this stage to
define $\mu _{ij}=h_{ij}+u_{i}u_{j}$, where $u^{i}$ is a unit
future-directed timelike vector on $\partial M$, associated with a foliation 
$\Sigma _{t}=\{t=$const$.\}$ of the spacetime $M$. Then $\hat{K}_{ij}=\frac{%
\mu _{ij}}{r}+\mathcal{O}(\rho ^{-(d-4)})$ and replacing it in (\ref{LGC})
one obtains: 
\begin{equation}
R_{ikjl}h^{kl}=-\frac{\Delta (h_{ij}+u_{i}u_{j})}{r}-\frac{d-4}{r}\Delta
_{ij}+\frac{\Delta _{il}u^{l}u_{j}+\Delta _{jl}u^{l}u_{i}}{r}.
\label{master}
\end{equation}%
In the following we shall denote $\tilde{\Delta}_{i}=\Delta _{ij}u^{j}$ and $%
\tilde{\Delta}=\Delta _{ij}u^{i}u^{j}$. More generally we shall indicate by
a tilded quantity the presence of at least one contraction of that quantity
with $u^{i}$.

Note that for $d=4$ one cannot find the general expression of $\Delta _{ij}$%
. However, as we shall subsequently show, when computing the action or the
conserved quantities one does not actually need $\Delta _{ij}$ but its
various contractions with the vector $u^{i}$. Now, contracting the above
relation with $u^{j}$ we obtain 
\begin{equation*}
\frac{d-3}{r}\tilde{\Delta}_{i}=-R_{ikjl}h^{kl}u^{j}+\frac{\tilde{\Delta}%
u_{i}}{r}
\end{equation*}%
and further contracting with $u^{i}$ we obtain

\begin{eqnarray}
\tilde{\Delta}&=&-\frac{r}{d-2} R_{ikjl}h^{kl}u^iu^j.
\end{eqnarray}

Recall that we are interested in Ricci-flat metrics and therefore we can
effectively replace in the above equations the curvature tensor in the bulk
with the bulk Weyl tensor. Using the earlier definition of the pull-back $%
E_{ij}$ to the boundary of the electric part of the bulk Weyl tensor we
obtain the following simple result

\begin{equation*}
\Delta =\frac{\tilde{\Delta}}{d-3}
\end{equation*}%
upon taking the trace of (\ref{master}) with $h^{ij}$.

In order to compute the boundary action we only have to evaluate the
difference $K-\hat{K}=h^{ij}\Delta _{ij}=\Delta $ and so we obtain 
\begin{eqnarray}
K-\hat{K} &=&-\frac{r}{(d-2)(d-3)}R_{ikjl}h^{kl}u^{i}u^{j}  \notag \\
&=&\frac{r}{(d-2)(d-3)}E_{ij}u^{i}u^{j}
\end{eqnarray}%
Finally, the regularized action is given by

\begin{eqnarray}
I_{\partial B}&=&-\frac{1}{8\pi G}\int_{\partial M}d^{d-1}x\sqrt{-h}\frac{r}{%
(d-2)(d-3)} E_{ij}u^iu^j.  \label{action}
\end{eqnarray}

Now, in order to evaluate the conserved charges we evaluate the boundary
stress-tensor using (see eq. (4.2) of \cite{Mann:2005yr})

\begin{equation*}
T_{ij}=\frac{1}{8\pi G}(\Delta _{ij}-h_{ij}\Delta ),
\end{equation*}%
from which the conserved charges associated with a boundary Killing vector $%
\xi $ are found by direct integration:

\begin{equation}
\mathcal{Q}=\oint_{\Sigma_t }d^{d-1}S^{i}T_{ij}\xi ^{j}.
\end{equation}

Recall that in $d=4$ we cannot directly evaluate $\Delta _{ij}$ so we cannot
find the explicit expression of $T_{ij}$. Fortunately to compute $\mathcal{Q}
$\ we only need the contraction of $T_{ij}$ with $u^{j}$

\begin{equation}
T_{ij}u^{j}=\frac{1}{8\pi G}(\tilde{\Delta}_{i}-\Delta u_{i})=\frac{1}{8\pi G%
}\frac{rE_{ij}u^{j}}{d-3},
\label{stress}
\end{equation}
such that the conserved charge associated with a Killing vector $\xi $ of
the boundary metric $h_{ij}$ is given by

\begin{equation}
\mathcal{Q}=\frac{1}{8\pi G}\oint_{\Sigma }d^{d-2}\sigma \frac{r E_{ij}
u^i\xi^{j}}{d-3}.  \label{conservedQ}
\end{equation}

Here, $\Sigma $ is a closed surface with unit normal $u^{i}$ and volume
element $d^{d-2}\sigma $. This completes our derivation of the regularized
action and  conserved quantities defined using the Mann-Marolf counterterm
in the cylindrical cut-off of the spacetime. For $d=4$ the relation (\ref%
{stress}) has been used previously in \cite{KKenergy} to compute the total
mass associated with the Kaluza-Klein magnetic monopole in four dimensions.

\section{Other counterterm prescriptions}

For asymptotically flat $4$-dimensional spacetimes the counterterm 
\begin{equation}
I_{ct}=\frac{1}{8\pi G}\int d^{3}x\sqrt{-h}\sqrt{2\mathcal{R}}
\label{Laumann}
\end{equation}%
was proposed \cite{Lau,Mann1} to eliminate divergences that occur in (\ref%
{actbulkgh}). An analysis of the higher dimensional case \cite{Kraus}
suggested in $d$-dimensions the counterterm:

\begin{equation}
I_{ct}=\frac{1}{8\pi G}\int d^{d-1}x\sqrt{-h}\frac{\mathcal{R}^{\frac{3}{2}}%
}{\sqrt{\mathcal{R}^{2}-\mathcal{R}_{ij}\mathcal{R}^{ij}}},  \label{Iflat1}
\end{equation}%
where $\mathcal{R}_{ij}$ is the Ricci tensor of the induced metric $h_{ij}$
and $\mathcal{R}$ is the corresponding Ricci scalar. This counterterm
removes divergencies in the action for a general class of asymptotically
flat spacetimes with boundary topologies $S^{n}\times R^{d-1-n}$. There also
exists another simpler counterterm that removes the divergencies in the
action for the general class of asymptotically flat spacetimes with boundary
topologies $S^{n}\times R^{d-1-n}$ \cite%
{Kraus,Mann:2005yr,Mann:2006bd,KKenergy,Hossein,Radu}:

\begin{equation}
I_{ct}=\frac{1}{8\pi G}\int d^{d-1}x\sqrt{-h}\sqrt{\frac{n\mathcal{R}}{n-1}}.
\label{Iflat2}
\end{equation}

By taking the variation of the total action (including the counterterm (\ref%
{Iflat2})) with respect to the boundary metric $h_{ij}$ the boundary
stress-tensor is given by \cite{Astefanesei:2005ad}

\begin{equation}
T_{ij}=\frac{1}{8\pi G}\left( K_{ij}-Kh_{ij}- \Psi( \mathcal{R}_{ij}-%
\mathcal{R}h_{ij})-h_{ij}\Box\Psi+\Psi_{;ij}\right),  \label{5dcounterterm}
\end{equation}
where we denoted $\Psi=\sqrt{\frac{n}{(n-1)\mathcal{R}}}$. If the boundary
geometry has an isometry generated by a Killing vector $\xi ^{i}$, then $%
T_{ij}\xi ^{j}$ is divergence free, from which it follows that the quantity 
\begin{equation}
\mathcal{Q}=\oint_{\Sigma }d^{d-2}\sigma T_{ij}u^i\xi ^{j},  \label{QsrtR}
\end{equation}
associated with a closed surface $\Sigma $, is conserved. Physically, this
means that a collection of observers on the boundary with the induced metric 
$h_{ij}$ measure the same value of $\mathcal{Q}$, provided the boundary has
an isometry generated by $\xi $. In particular, if $\xi ^{i}=\partial
/\partial t$ then $\mathcal{Q}$ is the conserved mass $\mathcal{M}$.

\section{Thermodynamics of NUT-charged spaces}

In this section we shall apply the results from the previous sections to
derive the action and the conserved charges associated with NUT-charged
spaces in four dimensions.

We start with the Euclidean form of the Taub-NUT solution \cite%
{Taub,NUT,Misner}: 
\begin{equation*}
ds^{2}=F_{E}(r)(d\chi -2n\cos \theta d\varphi
)^{2}+F_{E}^{-1}(r)dr^{2}+(r^{2}-n^{2})d\Omega ^{2},
\end{equation*}
where 
\begin{equation}
F_{E}(r)=\frac{r^{2}-2mr+n^{2}}{r^{2}-n^{2}}  \label{TN}
\end{equation}
In general, the $U(1)$ isometry generated by the Killing vector $\frac{%
\partial }{\partial \chi }$ (that corresponds to the coordinate $\chi $ that
parameterises the fibre $S^{1}$) can have a zero-dimensional fixed point set
(referred to as a `Nut' solution) or a two-dimensional fixed point set
(correspondingly referred to as a `Bolt' solution). The regularity of the
Euclidean Taub-Nut solution requires that the period of $\chi $ be $%
\beta=8\pi n$ (to ensure removal of the Dirac-Misner string singularity), $%
F_{E}(r=n)=0$ (to ensure that the fixed point of the Killing vector $\frac{%
\partial }{\partial \chi }$ is zero-dimensional) and also $\beta
F_{E}^{\prime }(r=n)=4\pi $ in order to avoid the presence of the conical
singularities at $r=n$. With these conditions we obtain $m=n$, yielding 
\begin{equation}
F_{E}(r)=\frac{r-n}{r+n}  \label{FEnut}
\end{equation}
The other possibility to explore is the Taub-Bolt solution in
four-dimensions \cite{Page}. In this case the Killing vector $\frac{\partial 
}{\partial \chi }$ has a two-dimensional fixed point set in the
four-dimensional Euclidean sector. The regularity of the solution is then
ensured by demanding that $r\geq 2n$, while the period of the coordinate $%
\chi $ is $8\pi n$ and for the bolt solution we obtain (with $m=5n/2$): 
\begin{equation}
F_{E}(r)=\frac{\left( r-2n\right) \left( r-\frac{1}{2}n\right) }{r^{2}-n^{2}}%
.  \label{FEbolt}
\end{equation}

Using the results derived using the Mann-Marolf prescription we obtain: 
\begin{eqnarray}
\Delta&=&-\frac{m}{r^2}+\mathcal{O}(r^{-3})
\end{eqnarray}
while to leading order the contracted boundary stress-tensor (\ref{stress})
has the following components: 
\begin{eqnarray}
8\pi GT_{\chi i}u^i&=&\frac{2m}{r^2}+\mathcal{O}(r^{-3}),  \notag \\
8\pi GT_{\phi i}u^i&=&\frac{6mn\cos\theta}{r^2}+\mathcal{O}(r^{-3}).
\label{MMnut}
\end{eqnarray}
Then the total mass and action are found to be: 
\begin{eqnarray}
\mathcal{M}&=&\frac{m}{G},~~~~~~I_{\partial B}=\frac{4\pi mn}{G}.
\end{eqnarray}

Using now the counterterm (\ref{Laumann}) one finds the following components
of the boundary stress-tensor \cite{Mann1}: 
\begin{eqnarray}
8\pi GT_{\chi }^{\chi } &=&\frac{2m}{r^{2}}+\mathcal{O}(r^{-3}),  \notag \\
8\pi GT_{\phi }^{\chi } &=&\frac{4mn\cos \theta }{r^{2}}+\mathcal{O}(r^{-3}),
\notag \\
8\pi GT_{\theta }^{\theta } &=&8\pi GT_{\phi }^{\phi }=\frac{2n^{2}-m^{2}}{%
2r^{3}}+\mathcal{O}(r^{-4}).
\label{Rnut}
\end{eqnarray}%
Therefore we find $\mathcal{M}=\frac{m}{G}$, while the action is given by $I=%
\frac{4\pi mn}{G}$. By making use of the quantum statistical relation the
entropy is simply given by: 
\begin{equation*}
S=\beta \mathcal{M}-I=\frac{4\pi mn}{G}.
\end{equation*}%
For the Nut solution $m=n$ while for the bolt solution $m=5n/2$ and our
results are consistent with those obtained previously \cite{Mann1}.

We can also compute directly the thermodynamic quantities and the action for
the Taub-NUT space in the Lorentzian section. The metric can be written as: 
\begin{equation*}
ds^{2}=-F(r)(dt -2N\cos \theta d\varphi
)^{2}+F^{-1}(r)dr^{2}+(r^{2}+N^{2})d\Omega ^{2},
\end{equation*}
where 
\begin{equation}
F(r)=\frac{r^{2}-2mr-N^{2}}{r^{2}+N^{2}}  \label{TNL}
\end{equation}
and in order to eliminate the Misner string singularities one has to
periodically identify the coordinate $t$ with period $\beta=8\pi N$. In the
Mann-Marolf prescription we obtain the mass and the action: 
\begin{eqnarray}
\mathcal{M}&=&\frac{m}{G},~~~~~~I_{\partial B}=\frac{4\pi mN}{G}.
\end{eqnarray}
Using now the counterterm (\ref{Laumann}) one finds the following components
of the boundary stress-tensor: 
\begin{eqnarray}
8\pi G T^{t}_{t}&=&\frac{2m}{r^2}+\mathcal{O}(r^{-3}),  \notag \\
8\pi G T^{t}_{\phi}&=&\frac{4mN\cos\theta}{r^2}+\mathcal{O}(r^{-3}),  \notag
\\
8\pi G T^{\theta}_{\theta}&=&8\pi G T^{\phi}_{\phi}=-\frac{2N^2+m^2}{2r^3}+%
\mathcal{O}(r^{-4}).
\end{eqnarray}
such that the conserved mass is $\mathcal{M}=m/G$ and the action is again
given by $I=4\pi mN/G$. The same results have been obtained in \cite%
{Mann:2004mi} by other means.

\section{Rotating black holes in four dimensions}

In this section we consider $4$-dimensional asymptotically flat rotating
black holes. Using the Mann-Marolf counterterm we shall study the
thermodynamic properties of the Kerr black hole and we shall compare our
results against the conserved charges and action as computed using the
counterterm (\ref{Laumann}).

To apply the counterterm-prescription it is more convenient to use the
expression of the Kerr metric in Boyer-Lindquist coordinates as in this case
there are no cross-terms between $dr$ and the other coordinate
differentials. This simplifies the analysis of the event horizons and the
causal structure of the metrics. Moreover Boyer-Lindquist coordinates are
valid either outside the horizon or inside the horizon. In these
coordinates, the Kerr metric is: 
\begin{eqnarray}
ds^{2} &=&-\frac{\left( \delta -a^{2}\sin ^{2}\theta \right) }{\sigma }%
dt^{2}-2a\sin ^{2}\theta \frac{\left( r^{2}+a^{2}-\delta \right) }{\sigma }%
dt\,d\phi  \\
&&+\left[ \frac{\left( r^{2}+a^{2}\right) ^{2}-\delta a^{2}\sin ^{2}\theta }{%
\sigma }\right] \sin ^{2}\theta d\phi ^{2}+\frac{\sigma }{\delta }%
dr^{2}+\sigma d\theta ^{2},
\end{eqnarray}%
where 
\begin{equation*}
\sigma =r^{2}+a^{2}\cos ^{2}\theta ,~~~~~\delta =r^{2}-2Mr+a^{2}.
\end{equation*}%
The two parameters describing the metric are $M$ and $a$.

Using the Mann-Marolf prescription we obtain from (\ref{stress}): 
\begin{eqnarray}
8\pi GT_{\chi i}u^{i} &=&\frac{2m}{r^{2}}+\mathcal{O}(r^{-3}),  \notag \\
8\pi GT_{\phi i}u^{i} &=&-\frac{3am\sin ^{2}\theta }{r^{2}}+\mathcal{O}(r^{-3}). 
 \label{MMKerr}
\end{eqnarray}
while $\Delta =-\frac{m}{r^{2}}+\mathcal{O}(r^{-3})$. We find then the total
mass $\mathcal{M}=m/G$ and the action $I_{\partial B}=\frac{\beta m}{2G}$,
where $\beta $ is the inverse temperature. It is worth  mentioning that the
stress-tensor can be computed on either the Lorentzian or the Euclidean
sections. However to compute the action we used the real Euclidean section
of the Kerr black hole. Notice that we can also compute the angular momentum 
$J$ using in (\ref{conservedQ}) the axial Killing vector $\xi =\frac{\partial }{\partial \phi }$. By simple integration, using the components listed in (\ref{MMKerr}), we find the angular momentum $J=-\frac{ma}{G}=-\mathcal{M}a$.

Using now the counterterm (\ref{Laumann}) we find the following components
of the boundary stress-tensor: 
\begin{eqnarray}
8\pi G T^{t}_{t}&=&\frac{2m}{r^2}+\mathcal{O}(r^{-3}),  \notag \\
8\pi G T^{t}_{\phi}&=&-\frac{3am\sin^2\theta}{r^2}+\mathcal{O}(r^{-3}), 
\notag \\
8\pi G T^{\phi}_{t}&=&\frac{3am}{r^4}+\mathcal{O}(r^{-5}),  \notag \\
8\pi G T^{\theta}_{\theta}&=&\frac{3a^2\cos^2\theta+a^2-m^2}{2r^3}+\mathcal{O%
}(r^{-4}),  \notag \\
8\pi G T^{\phi}_{\phi}&=&\frac{9a^2\cos^2\theta-5a^2-m^2}{2r^3}+\mathcal{O}%
(r^{-4}).
\end{eqnarray}

Also, using (\ref{QsrtR}) we find that the total mass is $\mathcal{M}=m/G$
while the total angular momentum is $J=-\mathcal{M}a$. Finally, a direct
computation reveals the total action to be $I={\beta \mathcal{M}}/2$, 
\textit{i.e.} the same results as the ones obtained by the Mann-Marolf
prescription. We also note that the conserved charges and the action are in
agreement with previous results in the literature (see for instance \cite{Hossein} and references therein).

\section{Conclusions}

The main motivation for this work was to obtain explicit expressions in the
so-called `cylindrical cut-off' for the boundary stress-tensor and the
renormalised action defined by using the boundary counterterm recently
proposed by Mann and Marolf for asymptotically flat spacetimes. In
particular, we showed that, even if in four dimensions one cannot find
directly the expression $\hat{K}_{ij}$ of the counterterm to first order,
one is still able to compute the conserved quantities and the renormalised
action. As with the results obtained previously in the `hyperbolic cut-off',
we found that the final action and the conserved charges in the `cylindrical
cut-offs' are essentially constructed using the electric part of the Weyl
tensor.

As an example of this technique we computed the action and the conserved
quantities for NUT-charged spaces in four dimensions (both in Euclidian and
Lorentzian sections) and for the Kerr solution. We showed explicitly our
results are consistent with that previously known in literature and overcome
difficulties inherent in the background subtraction approach \cite{Martinez}. In particular, we compared the results with the ones derived from a counterterm proportional with the square-root of the boundary Ricci scalar proposed by \cite{Lau,Mann1}.

In general, different counterterms can lead to different results when
computing the energy and the total action, seriously constraining the
various choices of the boundary counterterms (see for instance \cite{Hollands1,Hollands2} for a general study of the counterterm charges and a comparison with charges computed by other means in AdS context). While it is clear that the distinct choices of counterterms yield the same mass and
action for the asymptotically flat spaces considered here, some of the components of the boundary stress-energy tensors obtained using various counterterms have slightly
different coefficients (for example compare the second equations in (\ref{MMnut}) and (\ref{Rnut}) respectively; alternatively, see the stress-energy components computed for the Kaluza-Klein monopole in \cite{KKenergy} using different counterterms). These components are important if one considers for
instance the conserved charges associated with bubbles of nothing. At first
sight therefore, different counterterms might lead to different results for
such spaces.

While our discussion has focussed on a particular class of stationary
solutions in four dimensions, namely on the NUT-charged spaces and on the
Kerr solution, using the results from this work it is also possible to
investigate more general stationary backgrounds that are asymptotically
flat, such as general rotating black objects in higher dimensions. Work on
this is in progress and it will be reported elsewhere \cite{dumi}.

Finally, we believe that our results warrant further study of the
counterterm method in asymptotically flat spacetimes.

\medskip 

\medskip {\Large Acknowledgements}

DA was supported by the Department of Atomic Energy, Government of India and
the visitor programme of ICTP, Trieste. The work of RBM and CS was supported
by the Natural Sciences and Engineering Research Council of Canada.

\end{document}